\documentclass[aps,10pt,twocolumn,preprintnumbers,amsmath,amssymb,prl]{revtex4-1}

\usepackage{graphicx}
\usepackage{braket}
\usepackage[colorlinks,citecolor=blue,linkcolor=blue,urlcolor=blue,plainpages=false,pdfpagelabels]{hyperref}

\def\imagetop#1{#1}

\newcommand{\sigbar}{{\bar{\sigma}}}
\newcommand{\ssigma}{{\hat{\sigma}}}
\newcommand{\srho}{{\hat{\rho}}}
\DeclareMathOperator{\csch}{csch}

\begin{document}

\title{Fundamental limits to helical edge conductivity due to spin-phonon scattering}

\author{Solofo Groenendijk}
\author{Giacomo Dolcetto}
\author{Thomas~L.~Schmidt}

\affiliation{Physics and Materials Science Research Unit, University of Luxembourg, L-1511 Luxembourg}

\date{\today}

\begin{abstract}
We study the effect of electron-phonon interactions on the electrical conductance of a helical edge state of a two-dimensional topological insulator. We show that the edge deformation caused by bulk acoustic phonons modifies the spin texture of the edge state, and that the resulting spin-phonon coupling leads to inelastic backscattering which makes the transport diffusive.
Using a semiclassical Boltzmann equation we compute the electrical conductivity and show that it exhibits a metallic Bloch-Gr\"uneisen law. At temperatures on the order of the Debye temperature of the host material, spin-phonon scattering thus lowers the conductivity of the edge state drastically. Transport remains ballistic only for short enough edges, and in this case the correction to the quantized conductance vanishes as $\delta G \propto T^5$ at low temperatures. Relying only on parallel transport of the helical spin texture along the deformed edge, the coupling strength is determined by the host material's density and sound velocity. Our results impose fundamental limits for the finite-temperature conductivity of a helical edge channel.
\end{abstract}

\maketitle

The one-dimensional edge channels of two-dimensional topological insulators (2D TIs) have been studied in detail both theoretically and experimentally \cite{qi11,hasan10} ever since their experimental discovery ten years ago \cite{koenig07}. If the system is time-reversal invariant, electronic transport in these edge channels differs markedly from transport in conventional one-dimensional quantum systems~\cite{dolcetto16}. The reason is Kramers theorem, which prohibits elastic backscattering between counterpropagating spin-$1/2$ electrons in 2D TIs. As a consequence, the zero-bias conductance at zero temperature remains quantized even if the system is subject to disorder or interactions. To a certain extent, this has been confirmed by experiments, which have measured a conductance close to the predicted value of $G_0 = e^2/h$ in short edge channels \cite{roth09,knez11,du15,reis17}.

Nevertheless, Kramers theorem does not impede transitions between counterpropagating electrons with different energies, i.e., \emph{inelastic} backscattering. The latter always requires interactions, such as for instance electron-electron \cite{xu06,wu06,maciejko09,tanaka11,schmidt12,crepin12,orth13,vayrynen13}, or electron-phonon interactions \cite{budich12,garate13}. Over the past years, various scattering mechanisms have been proposed and each has been shown to cause a temperature-dependent correction to the edge conductance at finite temperature $T$ or finite voltages $V$, such that in general the conductance of a single edge is reduced to $G(T,V) = G_0 - \delta G(T,V)$.

Nevertheless, even at finite energies it remains true that the backscattering mechanisms which are most detrimental for the conductance of conventional 1D systems have a weaker effect in topological insulator edge channels. The reason is that, in addition to interactions, inelastic backscattering in edge channels requires a way to flip an electron's spin. There are various possibilities to flip spins even in a time-reversal invariant system, for instance Rashba spin-orbit coupling \cite{kane05,virtanen12,schmidt12,orth15,rod15,rod16}, which breaks the axial spin symmetry globally, or local spin-flipping perturbations such as Rashba scatterers~\cite{crepin12}, Kondo impurities~\cite{maciejko09,tanaka11,maciejko12}, or charge puddles~\cite{vayrynen13}.

In this paper, we will revisit the problem of electron-phonon scattering in a helical system and we will argue that electron-phonon coupling is a more important scattering mechanism than previously thought. This has two reasons: first, we will show that the lattice deformations caused by phonons, in tandem with the helical spin texture of the electrons, give rise to ``spin-phonon'' coupling. This coupling between the electrons' spins and the phonons alone can lead to backscattering, even without additional spin-flipping mechanisms such as impurities or Rashba spin-orbit coupling.

Moreover, we show that the coupling to \emph{transverse} phonons is essential. Whereas the electrons propagate in one-dimensional edge channels, the phonons exist in a three-dimensional crystal. Scattering with longitudinal phonons is kinematically suppressed because of the large difference between the sound velocity of acoustic phonons and the Fermi velocity of electrons. This problem is avoided for transverse phonons because the lattice vibrations perpendicular to the edge also carry elastic energy.

The structure of this paper is as follows. After motivating the existence of a spin-phonon coupling term from symmetry considerations, we will derive this term by using parallel transport of the electronic Dirac spectrum along a one-dimensional edge deformed by a phonon. For a short edge, we find a correction to the quantized conductance $\delta G(T)$ which exceeds previously known corrections originating from the interplay of Rashba impurities and scattering with longitudinal phonons \cite{budich12}. We will then move on to study the limit of long edges, where transport becomes diffusive due to spin-phonon coupling. We will calculate the resistivity of the edge state as a function of temperature using the Boltzmann equation and find that its temperature-dependence is given by a Bloch-Gr\"uneisen law.

We would like to point out that spin-phonon coupling makes the edge transport diffusive even in an ``ideal'' helical edge state which is free of impurities, Rashba spin-orbit coupling and electron-electron interactions. Since all these effects would only increase backscattering, our results provide a minimum resistivity for a helical edge channel at finite temperatures. Moreover, being independent of microscopic details, they are valid for all realizations of two-dimensional topological insulators. In particular, we expect this type of scattering to be a hindrance for potential low-dissipation electronic applications of 2D TIs. The required room-temperature 2D TIs have been proposed in several materials \cite{xu13,li15c,fu16,yuan17}, and have come a step closer with recent experiments on bismuthene \cite{reis17}.

To illustrate the main concepts, we begin by constructing a spin-phonon coupling Hamiltonian based solely on symmetry arguments. Consider the following toy Hamiltonian which couples the spin of the electrons to phonons,
\begin{align}\label{eq:toy}
H_1
= \sum_{k,q}q\left[ c_{k+q,\downarrow}^\dagger c_{k,\uparrow} \big(a^\dagger_{-q} - a_{q} \big) + \text{h.c.} \right],
\end{align}
where $q$ and $k$ denote electron and phonon momenta. We define the anti-unitary time-reversal operator $\Theta$ to have the usual effect on the fermionic annihilation operators, $\Theta c_{k,\uparrow}\Theta^{-1}=c_{-k,\downarrow}$ and $\Theta c_{k,\downarrow}\Theta^{-1}=-c_{-k,\uparrow}$. The minus sign in the last equation ensures that $\Theta^2=-1$ when acting on a state with an odd number of fermions. For the phonons time-reversal merely flips the momentum, $\Theta a_{q}\Theta^{-1}=a_{-q}$. Hamiltonian (\ref{eq:toy}) is time reversal invariant, $[H_1,\Theta]=0$, and it can cause inelastic backscattering between edge electrons. In a system with cylindrical symmetry, it can be regarded as a process in which electrons change their momentum and flip their spin by absorbing or emitting a phonon with angular momentum $\pm \hbar$ and linear momentum $q$ \cite{garanin15,zhang14a}.

\begin{figure}
\centering
\begin{tabular}{llll}
    (a) & \imagetop{\includegraphics[width=3.5cm]{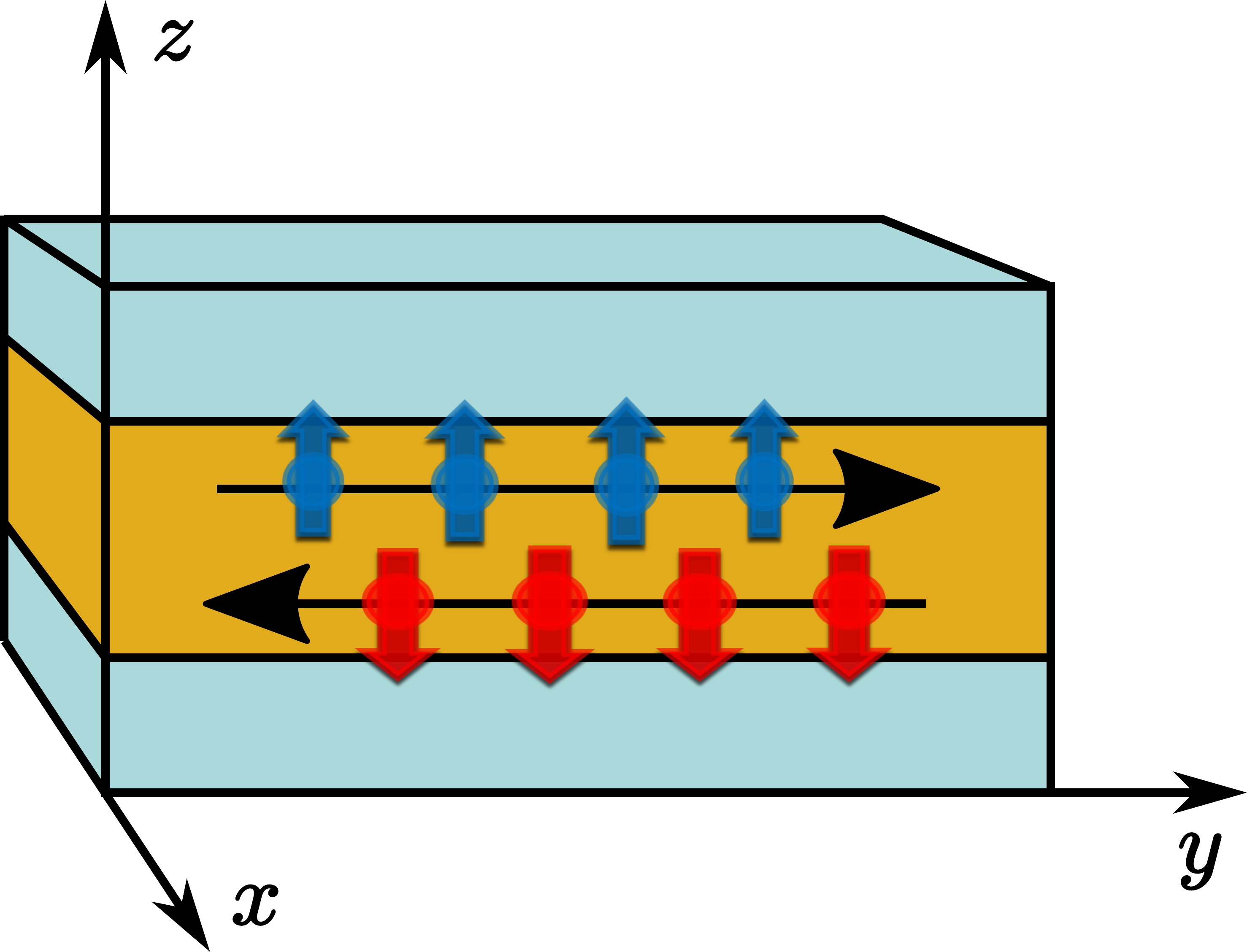} }&
    (b) & \imagetop{\includegraphics[width=3.5cm]{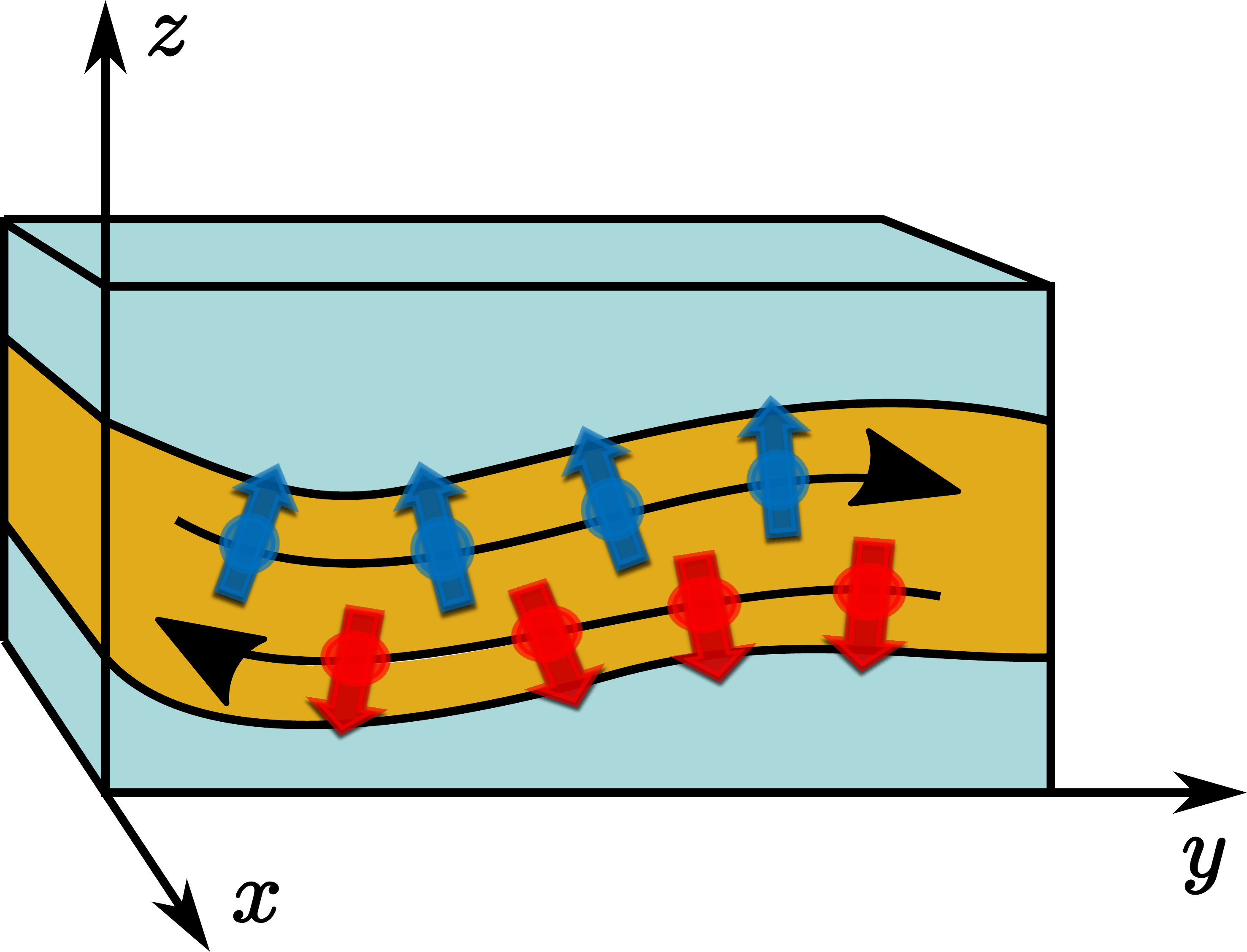} }
\end{tabular}
\caption{(a) A two-dimensional topological insulator in a semiconductor heterostructure, where we assume the spin quantization axis to be along the growth direction ($z$ direction). The electrons form a helical channel along the $y$ axis, where opposite spins propagate in opposite directions. (b) Due to spin momentum locking, small deformations of the edge lead to a position-dependent spin texture. \label{fig:sketch}}
\end{figure}

To see how an electron-phonon coupling term with the structure (\ref{eq:toy}) can emerge in a topological insulator edge state, we consider a 3D heterostructure grown along the $z$ direction as shown in Fig.~\ref{fig:sketch}. Its surface hosts a translationally invariant clean 1D helical edge channel in the $y$ direction. Moreover, we assume that the spin quantization axis is along the growth direction. Hence, we consider mainly systems where the axial spin symmetry is not broken, but below we also comment briefly on the case when it is broken due to, e.g., bulk inversion asymmetry. The edge channel resides between the bulk of the 2D TI material $(x<0)$ and vacuum $(x>0)$. The kinetic energy of the helical edge electrons is given by the following Hamiltonian: \cite{dolcetto16}
\begin{equation}
H_{\rm el} = v_F\int dy\ \Psi^\dag(y)\sigma_z\hat{p}_y\Psi(y).
\end{equation}
Here, $\hat{p}_y = -i\hbar\partial_y$ denotes the momentum operator along the $y$ direction, $v_F$ is the Fermi velocity, and $\sigma_z$ is a Pauli matrix in spin space. Moreover, $\Psi = (\psi_\uparrow,\psi_\downarrow)^T$ are spinors where $\psi_{\uparrow(\downarrow)}$ annihilates a right (left) moving spin-up (spin-down) electron.

Next, we consider a static deformation of the edge channel due to a displacement of the crystal ions from their equilibrium positions as shown in Fig.~\ref{fig:sketch}. For small long-wavelength distortions, spin-momentum locking imposes that the spins will remain perpendicular to the local propagation direction of the electrons. In Fig.~\ref{fig:sketch}, the spin will thus acquire a component along the $y$ direction. To model this process mathematically, we consider first the effect of a rotation by an angle $\phi_x$ about the $x$ axis on the edge electrons. For the spin, this corresponds to the unitary transformation $U(\phi_x) = \exp(i \phi_x \sigma_x/2)$, which for an infinitesimal rotation yields then $\sigma'_z= U^\dag(\phi_x)\sigma_zU(\phi_x) \approx \sigma_z - \phi_x \sigma_y$. The momentum operator transforms as $\hat{p}'_y\approx\hat{p}_y-\phi_x \hat{p}_z$. If we allow for a slow position dependence of the rotation angle $\phi_x(y)$, we thus find,
\begin{equation}
\sigma_z\hat{p}_y \to \sigma_z\hat{p}_y - \phi_x(y) \sigma_y\hat{p}_y- \phi_x(y) \sigma_z \hat{p}_z.
\end{equation}
The second term represents the position-dependent tilting of the spin quantization axis and contains $\sigma_y$, which is off-diagonal in spin space and causes spin flips. The third term represents the transformation of the momentum operator and is proportional to $\hat{p}_z$. Since this term is diagonal in spin space, it does not contribute significantly to backscattering effects and will from now on be discarded.

In order to make the static rotational deformation field $\phi_x(y)$ dynamical, we quantize it in terms of phonons. The ionic displacement field $\mathbf{u}(\mathbf{r})$ is related to the rotational field (also known as vorticity field) by $\boldsymbol{\phi} = \frac{1}{2} (\boldsymbol{\nabla}\times\mathbf{u})$~\cite{spencer04,garanin15}. By taking the curl of the quantized displacement field $\mathbf{u}(\mathbf{r})$ we thus end up with the following quantized rotational deformation field,
\begin{equation}
\boldsymbol{\phi}(\mathbf{r})= \frac{i}{\sqrt{\Omega}}\sum_{\mathbf{q},\lambda}\sqrt{\frac{\hbar}{8\rho\omega(\mathbf{q})}}(\mathbf{q}\times\boldsymbol{\xi}^\lambda_{\mathbf{q}}) e^{i\mathbf{q}\mathbf{r}}\big(a_{\mathbf{q},\lambda}+a^\dagger_{-\mathbf{q},\lambda}\big),
\end{equation}
where $\rho$ is the mass density of the crystal, and $\boldsymbol{\xi}^\lambda_{\mathbf{q}}$ is the transverse polarization vector of a phonon of momentum $\mathbf{q}$ and polarization $\lambda \in \{1,2\}$, which satisfies $\boldsymbol{\xi}^\lambda_{\mathbf{q}}\cdot\mathbf{q}=0$. Without loss of generality, we assume the polarization vectors to be real, so $\boldsymbol{\xi}^\lambda_{\mathbf{q}}={\boldsymbol{\xi}^\lambda_{-\mathbf{q}}}$. Finally, $\Omega$ is the volume of the topological insulator. The phonon Hamiltonian is $H_{\rm ph} = \sum_{\mathbf{q},\lambda}\hbar\omega(\mathbf{q})a^\dagger_{\mathbf{q},\lambda}a_{\mathbf{q},\lambda}$. We consider only acoustic phonons with linear dispersion $\omega(\mathbf{q})=c_s|\mathbf{q}|$ determined by the sound velocity $c_s$.

As a consequence of edge deformation, the Hamiltonian of the helical edge states thus changes as $H_{\rm el}\to H_{\rm el}+H_{\rm sp-ph}$, where the additional term is the following spin-phonon Hamiltonian,
\begin{equation}\label{eq:SpinPhonon}
H_{\rm sp-ph} = -\frac{v_F}{2}\int dy\ \phi_x(y) \Psi^\dag(y)\sigma_y \hat{p}_y\Psi(y) + \text{h.c.}
\end{equation}
where we used $\phi_x(y) \equiv \phi_x(x=0,y)$ because the phonon wavelength is large compared to the penetration depth of the electronic surface states into the bulk. The Hamiltonian (\ref{eq:SpinPhonon}) describes spin flips of the electrons caused by inelastic scattering with transversal phonons. Since the rotational deformation field is time reversal invariant, the same holds true for the spin-phonon Hamiltonian. In Fourier space, Eq.~(\ref{eq:SpinPhonon}) writes as:
\begin{align}
H_{\rm sp-ph}&= -\frac{v_F}{\sqrt{\Omega}} \sum_{k,\mathbf{q},\lambda}V_{\mathbf{q}}^\lambda (2k+q_{\parallel}) \notag \\
&\times \Big(c^\dagger_{k+q_\parallel,\uparrow}c_{k,\downarrow}-c^\dagger_{k+q_\parallel,\downarrow}c_{k,\uparrow}\Big)
\big(a_{\mathbf{q},\lambda}+a^\dagger_{-\mathbf{q},\lambda}\big)\nonumber.
\end{align}
where $c_{k,\sigma}= L^{-1/2}\sum_k  e^{iky}\psi_\sigma(y)$ and $L$ is the edge length. Here, $\mathbf{q}=(q_x,q_\parallel,q_z)$ denotes the phonon momentum and $k$ is the edge electron momentum along the $y$ direction. The scattering potential $V^\lambda_{\mathbf{q}}(k) = [32\hbar^{-1}\rho\omega(\mathbf{q})]^{-1/2}k[(\mathbf{q}\times\boldsymbol{\xi}^\lambda_{\mathbf{q}})\cdot\mathbf{e}_x]$ depends only on material parameters like the mass density and the sound velocity. It can be shown that $H_{\rm sp-ph}$ is time-reversal invariant.

We would like to stress that this spin-phonon coupling mechanism is generic in the sense that it emerges as a consequence of spin-momentum locking in a deformed edge state. Such deformations are generated by phonons and will thus always be present at finite temperatures, even in clean samples. Moreover, the spin flips are brought about by the deformation of the edge itself. 

If axial spin symmetry is broken, e.g., due to bulk or structural inversion asymmetry, the spin will generally not point along the growth direction as shown in Fig.~\ref{fig:sketch} \citep{qi11,dolcetto16,rod15,rod16,schmidt12,maciejko10}.
However, even in this case, the interplay of the helical spin texture with deformations due to phonons will give rise to the same spin-phonon coupling mechanism albeit with different phonon boundary conditions. Finally, let us point out that a similar spin-phonon scattering mechanism due to flexural phonons is known in graphene quantum dots \cite{struck10,rudner10,ochoa12,droth_phd}.

In the following, we will investigate how $H_{\rm sp-ph}$ affects the zero-bias conductance of the edge at finite temperatures. The full system Hamiltonian for small deformations reads
\begin{equation}
H =  H_{\rm el}+H_{\rm ph}+H_{\rm sp-ph},
\end{equation}
where $H_{\rm sp-ph}$ will be treated as the perturbation using the Kubo formula for short edges and the Boltzmann equation for long edges.

\paragraph{Conductance of short edges:} We start by considering a short helical edge channel connected to two electron reservoirs with equal temperature $T$ but different chemical potentials $\mu_R =\mu-eV/2$ and $\mu_L =\mu+eV/2$ respectively. In the absence of backscattering, the right-moving (left-moving) helical electrons remain in equilibrium with the left (right) reservoir throughout the entire sample. If backscattering is weak and the edge is short (ballistic regime), their distribution function is only slightly changed compared to the noninteracting limit \cite{lunde07,rech09,micklitz10,imambekov12,schmidt13b}. If we consider the linear-response regime in the voltage, we can compute the backscattering conductance using the Kubo formula for conductance.

\begin{figure}[t]
\includegraphics[width=\columnwidth]{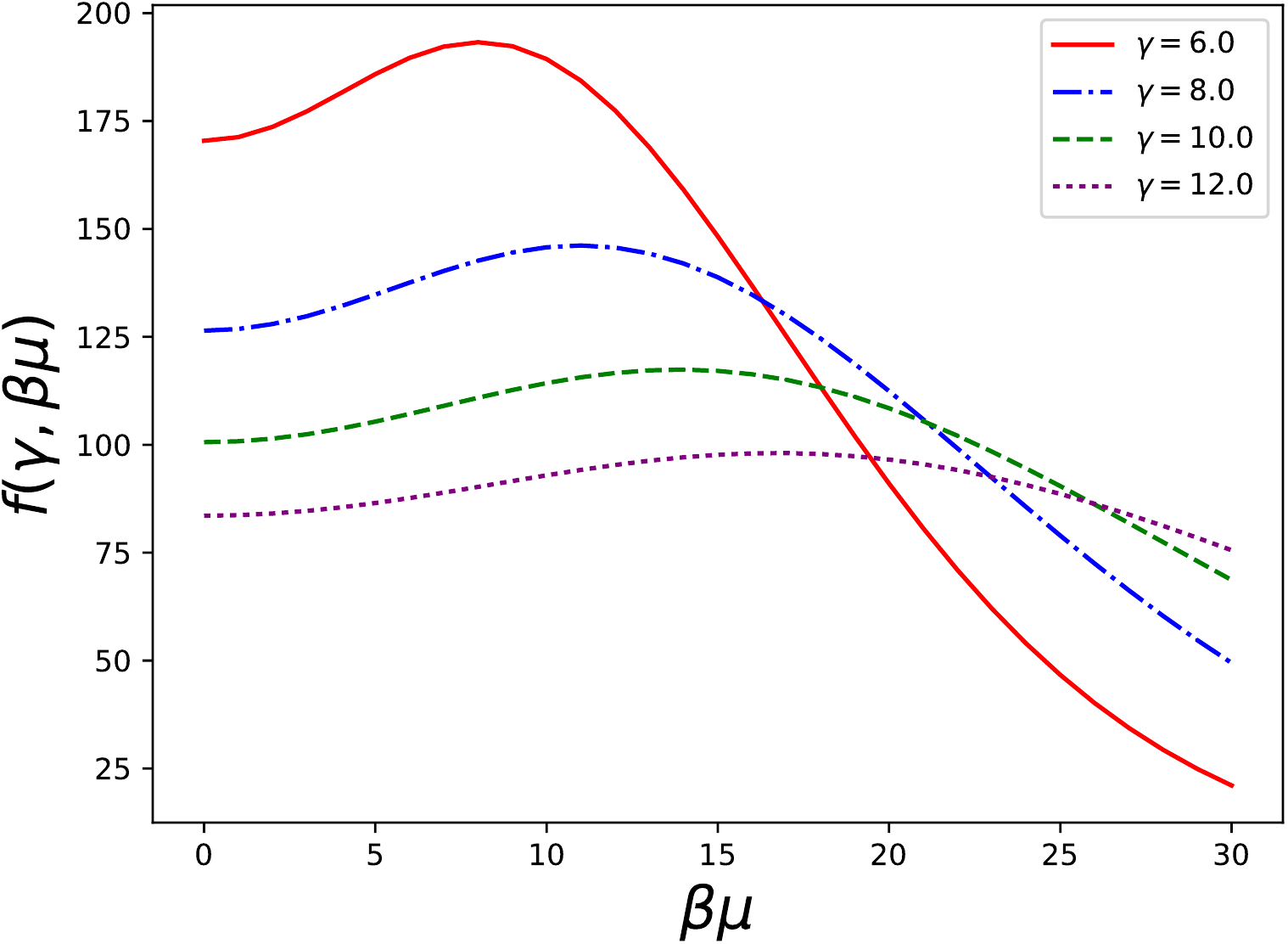}
\caption{The function $f(\gamma,\mu/T)$ for various values of $\gamma = v_F/c_s$.}
\label{fig:plotF}
\end{figure}

In the presence of electron-phonon scattering, the conductance is modified to
$\langle I\rangle\approx G_0V+\langle \delta I\rangle$. In the case of conserved total number of particles, we can define the backscattering current operator using the Heisenberg equation of motion as $\delta I=e (\dot{n}_\uparrow-\dot{n}_\downarrow)=2e [n_\uparrow,H_{\rm sp-ph}]/i\hbar$, where we defined the number operator $n_\sigma = \sum_k c^\dag_{k,\sigma}c_{k,\sigma}$.
The average current can be expressed as follows using the Kubo formula:
\begin{equation}\label{eq:kubo}
\left\langle\delta I(t)\right\rangle= \frac{1}{i\hbar}\int_{-\infty}^tdt^\prime\left\langle\left [\delta I(t),H_{\rm sp-ph}(t^\prime)\right ]\right\rangle_0,
\end{equation}
where $\left\langle\cdots\right\rangle_0$ denotes the expectation value with respect to the unperturbed ground state and the time dependence of the operators in the interaction picture is determined by the unperturbed Hamiltonian $H_{\rm el} + H_{\rm ph}$. For our Hamiltonian, $\left\langle \delta I(t) \right\rangle$ is time-independent, and a straightforward calculation yields the backscattering conductance,
\begin{align}\label{eq:fin}
\delta G &= - \frac{\left\langle\delta I\right\rangle}{V} = \frac{\hbar L G_0}{2^7\pi\rho v_F}\left(\frac{T}{\hbar c_s}\right)^5 f\left(\gamma, \mu/T\right),\\
f(\gamma,\mu/T) &=\int_0^{\infty} dx\int_{-1}^1d\lambda   \frac{x^5(\lambda^2+1)\csch\left(\tfrac{x}{2}\right)}{ \cosh\left(\tfrac{x}{2}\gamma\lambda+\frac{\mu}{T}\right) + \cosh\left(\tfrac{x}{2}\right)} \notag ,
\end{align}
where $\gamma = v_F/c_s \gg 1$. Here and in the following, we set the Boltzmann constant to unity.

If the chemical potential is at the Dirac point ($\mu = 0$), we find a reduction of the conductance, $G = G_0 - \delta G$ where $\delta G \propto T^5$. As shown in Fig.~\ref{fig:plotF}, for small $|\mu/T| \ll \gamma$, one finds $f(\gamma, \mu/T) - f(\gamma,0) \propto \mu/(T^2\gamma^{3}) $, which leads to an increase of $\delta G$ away from $\mu = 0$. For $|\mu/T| \ll \gamma$, this conductance correction thus always exceeds that obtained for longitudinal phonons~\cite{budich12}. On the contrary, for $|\mu/T| \gg \gamma$, the correction to the conductance becomes exponentially suppressed, $\delta G \propto e^{-\frac{\mu}{T}}$.

\begin{figure}
\includegraphics[width=\columnwidth]{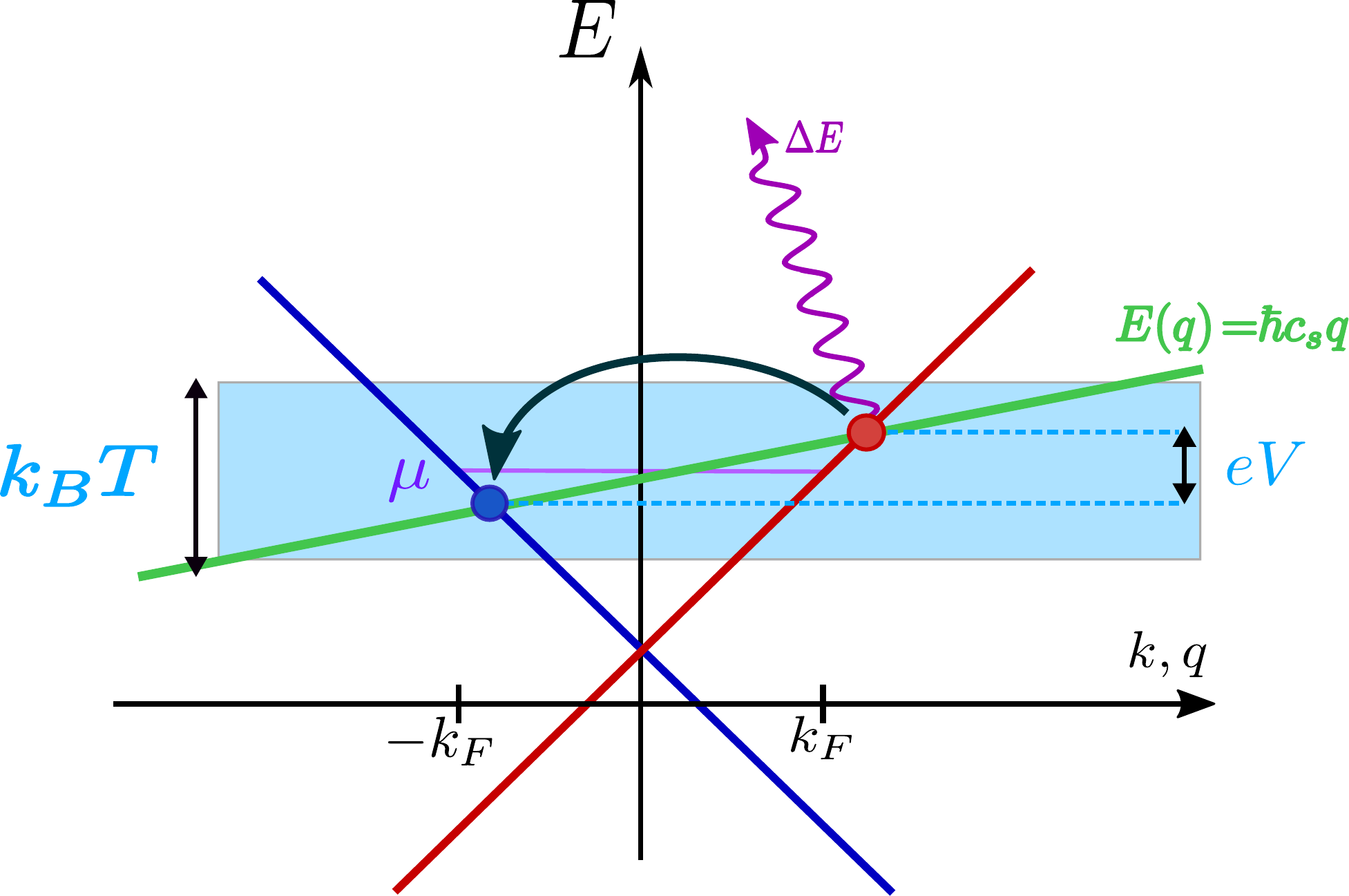}
\caption{A right-mover scatters into a left-mover while emitting a phonon (wavy purple line). The chemical potentials of right- and left-movers are shifted by the bias voltage $V$. In the linear-response limit $eV \ll T$, the temperature defines the energy window where this process is allowed.}
\label{fig:kinematics}
\end{figure}

Both features can be understood from Fig.~\ref{fig:kinematics}. Increasing the chemical potential from $\mu=0$ while keeping $\mu \ll \gamma T$ opens additional scattering channels, and thus increases $\delta G$. For $\mu \gg \gamma  T$, in contrast, the energy difference between right-moving initial state and left-moving final state becomes large. Therefore, the Fermi distributions require an increased temperature to allow filled right-moving and empty left-moving states.

\paragraph{Resistivity of long edges:} The conductance correction $\delta G$, which was obtained perturbatively in Eq.~(\ref{eq:fin}), grows linearly with the system length $L$ and is thus applicable only for short edges. The critical length $\ell_c$ can be defined as the length where $\delta G$ becomes of order $G_0$ \cite{rech09,micklitz10}, and the perturbative result breaks down. In the regime $L \gg \ell_c$, electrons flowing from one contact to the other undergo many scattering events, which allow them ultimately to relax to a quasi-equilibrium distribution inside the edge. Hence, the electrons lose the memory of the distribution function they originated from, and are driven by the electric field $E = V/L$ along the edge channel created by the bias difference \cite{rech09,micklitz10,imambekov12}. For long wires, electron transport becomes thus diffusive and the conductivity is given by Ohm's law as $\ssigma =  I/E$.

We will use a semi-classical Boltzmann equation to compute the conductivity. First we expand the distribution functions of the right- and left-movers with momentum $k$ to first order in the electric field as follows:
\begin{align}\label{eq:linearb}
n_\sigma(k) &= n^{(0)}_{\sigma}(k)+\delta n_{\sigma}(k),  \\
n^{(0)}_{\sigma}(k) &= \left( e^{ \frac{\varepsilon_\sigma(k)}{T}}+1 \right)^{-1}, \notag
\end{align}
where $\sigma =\ \uparrow,\downarrow\ = +,-$ labels the electron species, and $\smash{n^{(0)}_{\sigma}(k)}$ denotes the equilibrium Fermi distribution at chemical potential $\mu = \hbar v_F k_F$. The helical edge dispersion relation reads $\varepsilon_\sigma(k)=\hbar v_F(\sigma k-k_F)$. The correction $\delta n_{\sigma}(k)$ reflects the response of the electron distribution to the electric field. The edge current can be expressed in terms of these corrections,
\begin{equation}
I = -ev_F\sum_{\sigma} \sigma \int \frac{dk}{2\pi}\delta n_\sigma(k),
\end{equation}
as the corresponding contribution to the current due to the equilibrium distributions $n^{(0)}_{\sigma}(k)$ vanishes.

The distribution functions can be determined from the linearised Boltzmann equation \cite{ziman01},
\begin{align}\label{eq:boltz}
& -\frac{eE}{\hbar}\partial_k n_\sigma^{(0)}(k) = \mathcal{I}[n_\sigma(k)],  \\
&    \mathcal{I}[n_\sigma(k)] \label{eq:scattering}
=
    \sum_{\lambda,\vec{q}}\Bigg\{
      \Gamma^+_{\sigbar\to\sigma} n_{\sigbar} [1-n_\sigma]
    - \Gamma^-_{\sigma\to\sigbar} n_\sigma [1-n_{\sigbar}] \notag \\
&   + \Gamma^-_{\sigbar\to\sigma} n_{\sigbar}[1-n_\sigma]
    - \Gamma^+_{\sigma\to\sigbar} n_\sigma [1-n_{\sigbar}] \Bigg\}.
\end{align}
where $\mathcal{I}[n_\sigma(k)]$ is the collision integral. We defined $\sigbar = -\sigma$ and used $n_{\sigbar} :=n_{\sigbar}(k+q_\parallel)$, $n_\sigma:=n_\sigma(k)$. Moreover, $\Gamma^{\pm}_{\sigma\to\sigbar}$ denotes the backscattering rate for electrons from spin $\sigma$ to spin $\sigbar$ via emission (absorption) of a phonon. The transition rates are given by Fermi's golden rule and have the usual form $\Gamma = \frac{2\pi}{\hbar}|\bra{f}H_{\rm sp-ph}\ket{i}|^2\delta(E_f-E_i)$. For example, for the phonon emission process $\Gamma^+_{\sigbar\to\sigma}$, the initial state $\ket{i} = c^\dag_{k+q_{\parallel},\sigbar} \ket{FS;N_{\mathbf{q},\lambda}}$ contains one electron on top of the Fermi sea and a thermal number of photons $N_{\mathbf{q},\lambda}$ with momentum $q$ and polarization $\lambda$. This state has an energy $E_i = \varepsilon_\sigbar(k+q_\parallel)+N_{\mathbf{q},\lambda}\hbar\omega(\mathbf{q})$. The final state $\ket{f} = c^\dag_{k,\sigma} \ket{FS,N_{\mathbf{q},\lambda}+1}$ has the energy $E_f= \varepsilon_\sigma(k)+(N_{\mathbf{q},\lambda}+1)\hbar\omega(\mathbf{q})$, and contains an electron with opposite spin and the emitted photon. The other processes are defined analogously.

Assuming that the collision integral depends only weakly on energy allows us to express it in terms of a relaxation time, $\mathcal{I}[n_\sigma(k)]= -\delta n_\sigma(k)/\tau_\sigma(k)$. Then, the Boltzmann equation (\ref{eq:boltz}) can be solved for the distribution functions $\delta n_\sigma(k)$, and we find the conductivity

\begin{equation}\label{eq:conduct}
\ssigma(T) = -2\frac{e^2}{\hbar} v_F\int\frac{dk}{2\pi}\partial_k f_+^{(0)}(k)\tau_+(k).
\end{equation}

Here, we defined
\begin{align}
	\tau_+(k) &=  \frac{v_F \rho}{\pi \hbar c_s} \frac{1}{[\Delta_5(T)+k^2 \ \Delta_3(T)]}, \notag\\
    \Delta_n(T) &= \left( \frac{  T}{\hbar c_s} \right)^n \mathcal{J}_n(\Theta_D/T), \notag \\
    \mathcal{J}_n(y) &= \int_0^y dx \frac{x^ne^x}{(e^x-1)^2},
\end{align}
where $\Theta_D = \hbar c_s \sqrt[3]{6\pi^2\rho}$ is the Debye temperature for a 3D solid \cite{ziman01}. The appearance of the Bloch-Gr\"uneisen functions $\mathcal{J}_n(y)$ \cite{ziman01} can be traced back to the bosonic occupation numbers in the scattering integral. The result for the resistivity $\srho = 1/\ssigma$ can easily be determined by numerically calculating the integral in Eq.~(\ref{eq:conduct}). In the following, we discuss the relevant limiting cases. In the limit $T \ll \Theta_D$ the Bloch-Gr\"uneisen integrals $\mathcal{J}_{3,5}$ become constants. For $\mu \gg  T$, we find,
\begin{equation}
\srho\left(T \ll \mu, \Theta_D\right) \approx \frac{3\zeta(3)}{16} \frac{\mu^2 c_s}{e^2 v_F^4 \rho} \left( \frac{ T}{\hbar c_s} \right)^3.
\end{equation}
where $\zeta(x)$ denotes the Riemann zeta function. On the other hand, for $\mu \ll T$ the result reads
\begin{equation}
\srho\left( \mu \ll T \ll \Theta_D\right) \approx \frac{64\zeta(5)}{15 } \frac{\hbar^2 c_s}{e^2 v_F^2 \rho} \left( \frac{ T}{\hbar c_s} \right)^5.
\end{equation}
In contrast, for $T \gg \Theta_D$, the result will depend on the Debye temperature. In the range where $\mu \ll \Theta_D\ll T$ one has,
\begin{equation}\label{eq:rho3}
\srho\left( \mu \ll \Theta_D \ll T\right) \approx \frac{1}{512}\frac{\hbar^2 c_s}{ e^2 v_F^2 \rho }\left( \frac{ \Theta _D}{\hbar c_s}\right)^5 \frac{T}{\Theta_D}.
\end{equation}
Interestingly, this means that for small $\mu$, our model predicts a metallic Bloch-Gr\"uneisen behavior for helical edge states. Indeed, it is known that for 3D metals, the phonon contribution to the resistivity scales as $T^5$ below the Debye temperature and linear in $T$ above the Debye temperature \cite{ziman01}.

\paragraph{Estimates for the parameters:}
Typical parameters of helical edge states realized in semiconductor structures \cite{koenig07,roth09,knez11,du15} are $c_s \approx 3\times 10^3$~m/s, $\rho \approx 7\times 10^3$~kg/m$^3$, $v_F \approx 10^5$~m/s, $\Theta_D \approx 250$~K\cite{handbook05}. Since we are interested in wide band gap 2DTI materials with possible room temperature application \cite{reis17,xu13,li15c,fu16,yuan17}, we take $T=295$~K. For these parameters Eq.~(\ref{eq:fin}) predicts a critical length for ballistic transport of $\ell_c \approx 1 \mu$m. For longer edges, transport becomes diffusive and Eq.~(\ref{eq:rho3}) predicts a resistivity $\srho \approx 7\times 10^6$~$\Omega/$m. This resistivity is comparable to that of a copper nanowire with a diameter of $15$~nm, $\srho_{NW} \approx 4\times10^6$~$\Omega/$m at room temperature \cite{aveek06}, which implies that 2DTI edge states are not necessarily suitable for room-temperature nanoelectronics applications. However, our results provide guidance on how to minimize this dissipation, e.g., by choosing materials with large densities or sound velocities.

To summarize, we have proposed a mechanism by which edge deformations caused by phonons can lead to inelastic backscattering of helical electrons in the edge states of a two-dimensional topological insulator. We derived the effective coupling Hamiltonian by using parallel transport of the electronic spin texture along a distorted edge. Due to this geometric origin, the coupling strength is determined solely by the sound velocity and the mass density of the topological insulator material. We investigated the resulting correction to the conductance for short and long edges, and discussed its temperature dependence. We would like to point out that the proposed spin-phonon coupling occurs even in clean edge states. Therefore, the resistivity we calculated can be interpreted as the fundamental minimum resistivity of an ideal edge state.

\begin{acknowledgments}
The authors acknowledge support by the Fonds National de la Recherche Luxembourg under grant ATTRACT 7556175.
\end{acknowledgments}

\bibliography{references}

\begin{thebibliography}{44}%
\makeatletter
\providecommand \@ifxundefined [1]{%
 \@ifx{#1\undefined}
}%
\providecommand \@ifnum [1]{%
 \ifnum #1\expandafter \@firstoftwo
 \else \expandafter \@secondoftwo
 \fi
}%
\providecommand \@ifx [1]{%
 \ifx #1\expandafter \@firstoftwo
 \else \expandafter \@secondoftwo
 \fi
}%
\providecommand \natexlab [1]{#1}%
\providecommand \enquote  [1]{``#1''}%
\providecommand \bibnamefont  [1]{#1}%
\providecommand \bibfnamefont [1]{#1}%
\providecommand \citenamefont [1]{#1}%
\providecommand \href@noop [0]{\@secondoftwo}%
\providecommand \href [0]{\begingroup \@sanitize@url \@href}%
\providecommand \@href[1]{\@@startlink{#1}\@@href}%
\providecommand \@@href[1]{\endgroup#1\@@endlink}%
\providecommand \@sanitize@url [0]{\catcode `\\12\catcode `\$12\catcode
  `\&12\catcode `\#12\catcode `\^12\catcode `\_12\catcode `\%12\relax}%
\providecommand \@@startlink[1]{}%
\providecommand \@@endlink[0]{}%
\providecommand \url  [0]{\begingroup\@sanitize@url \@url }%
\providecommand \@url [1]{\endgroup\@href {#1}{\urlprefix }}%
\providecommand \urlprefix  [0]{URL }%
\providecommand \Eprint [0]{\href }%
\providecommand \doibase [0]{http://dx.doi.org/}%
\providecommand \selectlanguage [0]{\@gobble}%
\providecommand \bibinfo  [0]{\@secondoftwo}%
\providecommand \bibfield  [0]{\@secondoftwo}%
\providecommand \translation [1]{[#1]}%
\providecommand \BibitemOpen [0]{}%
\providecommand \bibitemStop [0]{}%
\providecommand \bibitemNoStop [0]{.\EOS\space}%
\providecommand \EOS [0]{\spacefactor3000\relax}%
\providecommand \BibitemShut  [1]{\csname bibitem#1\endcsname}%
\let\auto@bib@innerbib\@empty
\bibitem [{\citenamefont {Qi}\ and\ \citenamefont {Zhang}(2011)}]{qi11}%
  \BibitemOpen
  \bibfield  {author} {\bibinfo {author} {\bibfnamefont {X.}~\bibnamefont
  {Qi}}\ and\ \bibinfo {author} {\bibfnamefont {S.}~\bibnamefont {Zhang}},\
  }\href {\doibase 10.1103/RevModPhys.83.1057} {\bibfield  {journal} {\bibinfo
  {journal} {Rev. Mod. Phys.}\ }\textbf {\bibinfo {volume} {83}},\ \bibinfo
  {pages} {1057} (\bibinfo {year} {2011})}\BibitemShut {NoStop}%
\bibitem [{\citenamefont {Hasan}\ and\ \citenamefont {Kane}(2010)}]{hasan10}%
  \BibitemOpen
  \bibfield  {author} {\bibinfo {author} {\bibfnamefont {M.~Z.}\ \bibnamefont
  {Hasan}}\ and\ \bibinfo {author} {\bibfnamefont {C.~L.}\ \bibnamefont
  {Kane}},\ }\href {\doibase 10.1103/RevModPhys.82.3045} {\bibfield  {journal}
  {\bibinfo  {journal} {Rev. Mod. Phys.}\ }\textbf {\bibinfo {volume} {82}},\
  \bibinfo {pages} {3045} (\bibinfo {year} {2010})}\BibitemShut {NoStop}%
\bibitem [{\citenamefont {K\"onig}\ \emph {et~al.}(2007)\citenamefont
  {K\"onig}, \citenamefont {Wiedmann}, \citenamefont {Br\"une}, \citenamefont
  {Roth}, \citenamefont {Buhmann}, \citenamefont {Molenkamp}, \citenamefont
  {Qi},\ and\ \citenamefont {Zhang}}]{koenig07}%
  \BibitemOpen
  \bibfield  {author} {\bibinfo {author} {\bibfnamefont {M.}~\bibnamefont
  {K\"onig}}, \bibinfo {author} {\bibfnamefont {S.}~\bibnamefont {Wiedmann}},
  \bibinfo {author} {\bibfnamefont {C.}~\bibnamefont {Br\"une}}, \bibinfo
  {author} {\bibfnamefont {A.}~\bibnamefont {Roth}}, \bibinfo {author}
  {\bibfnamefont {H.}~\bibnamefont {Buhmann}}, \bibinfo {author} {\bibfnamefont
  {L.~W.}\ \bibnamefont {Molenkamp}}, \bibinfo {author} {\bibfnamefont {X.-L.}\
  \bibnamefont {Qi}}, \ and\ \bibinfo {author} {\bibfnamefont {S.-C.}\
  \bibnamefont {Zhang}},\ }\href {\doibase 10.1126/science.1148047} {\bibfield
  {journal} {\bibinfo  {journal} {Science}\ }\textbf {\bibinfo {volume}
  {318}},\ \bibinfo {pages} {766} (\bibinfo {year} {2007})}\BibitemShut
  {NoStop}%
\bibitem [{\citenamefont {Dolcetto}\ \emph {et~al.}(2016)\citenamefont
  {Dolcetto}, \citenamefont {Sassetti},\ and\ \citenamefont
  {Schmidt}}]{dolcetto16}%
  \BibitemOpen
  \bibfield  {author} {\bibinfo {author} {\bibfnamefont {G.}~\bibnamefont
  {Dolcetto}}, \bibinfo {author} {\bibfnamefont {M.}~\bibnamefont {Sassetti}},
  \ and\ \bibinfo {author} {\bibfnamefont {T.~L.}\ \bibnamefont {Schmidt}},\
  }\href {\doibase 10.1393/ncr/i2016-10121-7} {\bibfield  {journal} {\bibinfo
  {journal} {Rivista del Nuovo Cimento}\ }\textbf {\bibinfo {volume} {39}},\
  \bibinfo {pages} {113} (\bibinfo {year} {2016})}\BibitemShut {NoStop}%
\bibitem [{\citenamefont {Roth}\ \emph {et~al.}(2009)\citenamefont {Roth},
  \citenamefont {Br\"une}, \citenamefont {Buhmann}, \citenamefont {Molenkamp},
  \citenamefont {Maciejko}, \citenamefont {Qi},\ and\ \citenamefont
  {Zhang}}]{roth09}%
  \BibitemOpen
  \bibfield  {author} {\bibinfo {author} {\bibfnamefont {A.}~\bibnamefont
  {Roth}}, \bibinfo {author} {\bibfnamefont {C.}~\bibnamefont {Br\"une}},
  \bibinfo {author} {\bibfnamefont {H.}~\bibnamefont {Buhmann}}, \bibinfo
  {author} {\bibfnamefont {L.~W.}\ \bibnamefont {Molenkamp}}, \bibinfo {author}
  {\bibfnamefont {J.}~\bibnamefont {Maciejko}}, \bibinfo {author}
  {\bibfnamefont {X.}~\bibnamefont {Qi}}, \ and\ \bibinfo {author}
  {\bibfnamefont {S.}~\bibnamefont {Zhang}},\ }\href {\doibase
  10.1126/science.1174736} {\bibfield  {journal} {\bibinfo  {journal}
  {Science}\ }\textbf {\bibinfo {volume} {325}},\ \bibinfo {pages} {294}
  (\bibinfo {year} {2009})}\BibitemShut {NoStop}%
\bibitem [{\citenamefont {Knez}\ \emph {et~al.}(2011)\citenamefont {Knez},
  \citenamefont {Du},\ and\ \citenamefont {Sullivan}}]{knez11}%
  \BibitemOpen
  \bibfield  {author} {\bibinfo {author} {\bibfnamefont {I.}~\bibnamefont
  {Knez}}, \bibinfo {author} {\bibfnamefont {R.-R.}\ \bibnamefont {Du}}, \ and\
  \bibinfo {author} {\bibfnamefont {G.}~\bibnamefont {Sullivan}},\ }\href
  {\doibase 10.1103/PhysRevLett.107.136603} {\bibfield  {journal} {\bibinfo
  {journal} {Phys. Rev. Lett.}\ }\textbf {\bibinfo {volume} {107}},\ \bibinfo
  {pages} {136603} (\bibinfo {year} {2011})}\BibitemShut {NoStop}%
\bibitem [{\citenamefont {Du}\ \emph {et~al.}(2015)\citenamefont {Du},
  \citenamefont {Knez}, \citenamefont {Sullivan},\ and\ \citenamefont
  {Du}}]{du15}%
  \BibitemOpen
  \bibfield  {author} {\bibinfo {author} {\bibfnamefont {L.}~\bibnamefont
  {Du}}, \bibinfo {author} {\bibfnamefont {I.}~\bibnamefont {Knez}}, \bibinfo
  {author} {\bibfnamefont {G.}~\bibnamefont {Sullivan}}, \ and\ \bibinfo
  {author} {\bibfnamefont {R.-R.}\ \bibnamefont {Du}},\ }\href {\doibase
  10.1103/PhysRevLett.114.096802} {\bibfield  {journal} {\bibinfo  {journal}
  {Phys. Rev. Lett.}\ }\textbf {\bibinfo {volume} {114}},\ \bibinfo {pages}
  {096802} (\bibinfo {year} {2015})}\BibitemShut {NoStop}%
\bibitem [{\citenamefont {Reis}\ \emph {et~al.}(2017)\citenamefont {Reis},
  \citenamefont {Li}, \citenamefont {Dudy}, \citenamefont {Bauernfeind},
  \citenamefont {Glass}, \citenamefont {Hanke}, \citenamefont {Thomale},
  \citenamefont {Sch\"afer},\ and\ \citenamefont {Claessen}}]{reis17}%
  \BibitemOpen
  \bibfield  {author} {\bibinfo {author} {\bibfnamefont {F.}~\bibnamefont
  {Reis}}, \bibinfo {author} {\bibfnamefont {G.}~\bibnamefont {Li}}, \bibinfo
  {author} {\bibfnamefont {L.}~\bibnamefont {Dudy}}, \bibinfo {author}
  {\bibfnamefont {M.}~\bibnamefont {Bauernfeind}}, \bibinfo {author}
  {\bibfnamefont {S.}~\bibnamefont {Glass}}, \bibinfo {author} {\bibfnamefont
  {W.}~\bibnamefont {Hanke}}, \bibinfo {author} {\bibfnamefont
  {R.}~\bibnamefont {Thomale}}, \bibinfo {author} {\bibfnamefont
  {J.}~\bibnamefont {Sch\"afer}}, \ and\ \bibinfo {author} {\bibfnamefont
  {R.}~\bibnamefont {Claessen}},\ }\href {\doibase 10.1126/science.aai8142}
  {\bibfield  {journal} {\bibinfo  {journal} {Science}\ }\textbf {\bibinfo
  {volume} {357}},\ \bibinfo {pages} {287} (\bibinfo {year}
  {2017})}\BibitemShut {NoStop}%
\bibitem [{\citenamefont {Xu}\ and\ \citenamefont {Moore}(2006)}]{xu06}%
  \BibitemOpen
  \bibfield  {author} {\bibinfo {author} {\bibfnamefont {C.}~\bibnamefont
  {Xu}}\ and\ \bibinfo {author} {\bibfnamefont {J.~E.}\ \bibnamefont {Moore}},\
  }\href {\doibase 10.1103/PhysRevB.73.045322} {\bibfield  {journal} {\bibinfo
  {journal} {Phys. Rev. B}\ }\textbf {\bibinfo {volume} {73}},\ \bibinfo
  {pages} {045322} (\bibinfo {year} {2006})}\BibitemShut {NoStop}%
\bibitem [{\citenamefont {Wu}\ \emph {et~al.}(2006)\citenamefont {Wu},
  \citenamefont {Bernevig},\ and\ \citenamefont {Zhang}}]{wu06}%
  \BibitemOpen
  \bibfield  {author} {\bibinfo {author} {\bibfnamefont {C.}~\bibnamefont
  {Wu}}, \bibinfo {author} {\bibfnamefont {B.~A.}\ \bibnamefont {Bernevig}}, \
  and\ \bibinfo {author} {\bibfnamefont {S.}~\bibnamefont {Zhang}},\ }\href
  {\doibase 10.1103/PhysRevLett.96.106401} {\bibfield  {journal} {\bibinfo
  {journal} {Phys. Rev. Lett.}\ }\textbf {\bibinfo {volume} {96}},\ \bibinfo
  {pages} {106401} (\bibinfo {year} {2006})}\BibitemShut {NoStop}%
\bibitem [{\citenamefont {Maciejko}\ \emph {et~al.}(2009)\citenamefont
  {Maciejko}, \citenamefont {Liu}, \citenamefont {Oreg}, \citenamefont {Qi},
  \citenamefont {Wu},\ and\ \citenamefont {Zhang}}]{maciejko09}%
  \BibitemOpen
  \bibfield  {author} {\bibinfo {author} {\bibfnamefont {J.}~\bibnamefont
  {Maciejko}}, \bibinfo {author} {\bibfnamefont {C.}~\bibnamefont {Liu}},
  \bibinfo {author} {\bibfnamefont {Y.}~\bibnamefont {Oreg}}, \bibinfo {author}
  {\bibfnamefont {X.}~\bibnamefont {Qi}}, \bibinfo {author} {\bibfnamefont
  {C.}~\bibnamefont {Wu}}, \ and\ \bibinfo {author} {\bibfnamefont
  {S.}~\bibnamefont {Zhang}},\ }\href {\doibase 10.1103/PhysRevLett.102.256803}
  {\bibfield  {journal} {\bibinfo  {journal} {Phys. Rev. Lett.}\ }\textbf
  {\bibinfo {volume} {102}},\ \bibinfo {pages} {256803} (\bibinfo {year}
  {2009})}\BibitemShut {NoStop}%
\bibitem [{\citenamefont {Tanaka}\ \emph {et~al.}(2011)\citenamefont {Tanaka},
  \citenamefont {Furusaki},\ and\ \citenamefont {Matveev}}]{tanaka11}%
  \BibitemOpen
  \bibfield  {author} {\bibinfo {author} {\bibfnamefont {Y.}~\bibnamefont
  {Tanaka}}, \bibinfo {author} {\bibfnamefont {A.}~\bibnamefont {Furusaki}}, \
  and\ \bibinfo {author} {\bibfnamefont {K.~A.}\ \bibnamefont {Matveev}},\
  }\href {\doibase 10.1103/PhysRevLett.106.236402} {\bibfield  {journal}
  {\bibinfo  {journal} {Phys. Rev. Lett.}\ }\textbf {\bibinfo {volume} {106}},\
  \bibinfo {pages} {236402} (\bibinfo {year} {2011})}\BibitemShut {NoStop}%
\bibitem [{\citenamefont {Schmidt}\ \emph {et~al.}(2012)\citenamefont
  {Schmidt}, \citenamefont {Rachel}, \citenamefont {von Oppen},\ and\
  \citenamefont {Glazman}}]{schmidt12}%
  \BibitemOpen
  \bibfield  {author} {\bibinfo {author} {\bibfnamefont {T.~L.}\ \bibnamefont
  {Schmidt}}, \bibinfo {author} {\bibfnamefont {S.}~\bibnamefont {Rachel}},
  \bibinfo {author} {\bibfnamefont {F.}~\bibnamefont {von Oppen}}, \ and\
  \bibinfo {author} {\bibfnamefont {L.~I.}\ \bibnamefont {Glazman}},\ }\href
  {\doibase 10.1103/PhysRevLett.108.156402} {\bibfield  {journal} {\bibinfo
  {journal} {Phys. Rev. Lett.}\ }\textbf {\bibinfo {volume} {108}},\ \bibinfo
  {pages} {156402} (\bibinfo {year} {2012})}\BibitemShut {NoStop}%
\bibitem [{\citenamefont {Cr\'epin}\ \emph {et~al.}(2012)\citenamefont
  {Cr\'epin}, \citenamefont {Budich}, \citenamefont {Dolcini}, \citenamefont
  {Recher},\ and\ \citenamefont {Trauzettel}}]{crepin12}%
  \BibitemOpen
  \bibfield  {author} {\bibinfo {author} {\bibfnamefont {F.}~\bibnamefont
  {Cr\'epin}}, \bibinfo {author} {\bibfnamefont {J.~C.}\ \bibnamefont
  {Budich}}, \bibinfo {author} {\bibfnamefont {F.}~\bibnamefont {Dolcini}},
  \bibinfo {author} {\bibfnamefont {P.}~\bibnamefont {Recher}}, \ and\ \bibinfo
  {author} {\bibfnamefont {B.}~\bibnamefont {Trauzettel}},\ }\href {\doibase
  10.1103/PhysRevB.86.121106} {\bibfield  {journal} {\bibinfo  {journal} {Phys.
  Rev. B}\ }\textbf {\bibinfo {volume} {86}},\ \bibinfo {pages} {121106}
  (\bibinfo {year} {2012})}\BibitemShut {NoStop}%
\bibitem [{\citenamefont {Orth}\ \emph {et~al.}(2013)\citenamefont {Orth},
  \citenamefont {Str\"ubi},\ and\ \citenamefont {Schmidt}}]{orth13}%
  \BibitemOpen
  \bibfield  {author} {\bibinfo {author} {\bibfnamefont {C.~P.}\ \bibnamefont
  {Orth}}, \bibinfo {author} {\bibfnamefont {G.}~\bibnamefont {Str\"ubi}}, \
  and\ \bibinfo {author} {\bibfnamefont {T.~L.}\ \bibnamefont {Schmidt}},\
  }\href {\doibase 10.1103/PhysRevB.88.165315} {\bibfield  {journal} {\bibinfo
  {journal} {Phys. Rev. B}\ }\textbf {\bibinfo {volume} {88}},\ \bibinfo
  {pages} {165315} (\bibinfo {year} {2013})}\BibitemShut {NoStop}%
\bibitem [{\citenamefont {V\"ayrynen}\ \emph {et~al.}(2013)\citenamefont
  {V\"ayrynen}, \citenamefont {Goldstein},\ and\ \citenamefont
  {Glazman}}]{vayrynen13}%
  \BibitemOpen
  \bibfield  {author} {\bibinfo {author} {\bibfnamefont {J.~I.}\ \bibnamefont
  {V\"ayrynen}}, \bibinfo {author} {\bibfnamefont {M.}~\bibnamefont
  {Goldstein}}, \ and\ \bibinfo {author} {\bibfnamefont {L.~I.}\ \bibnamefont
  {Glazman}},\ }\href {\doibase 10.1103/PhysRevLett.110.216402} {\bibfield
  {journal} {\bibinfo  {journal} {Phys. Rev. Lett.}\ }\textbf {\bibinfo
  {volume} {110}},\ \bibinfo {pages} {216402} (\bibinfo {year}
  {2013})}\BibitemShut {NoStop}%
\bibitem [{\citenamefont {Budich}\ \emph {et~al.}(2012)\citenamefont {Budich},
  \citenamefont {Dolcini}, \citenamefont {Recher},\ and\ \citenamefont
  {Trauzettel}}]{budich12}%
  \BibitemOpen
  \bibfield  {author} {\bibinfo {author} {\bibfnamefont {J.~C.}\ \bibnamefont
  {Budich}}, \bibinfo {author} {\bibfnamefont {F.}~\bibnamefont {Dolcini}},
  \bibinfo {author} {\bibfnamefont {P.}~\bibnamefont {Recher}}, \ and\ \bibinfo
  {author} {\bibfnamefont {B.}~\bibnamefont {Trauzettel}},\ }\href {\doibase
  10.1103/PhysRevLett.108.086602} {\bibfield  {journal} {\bibinfo  {journal}
  {Phys. Rev. Lett.}\ }\textbf {\bibinfo {volume} {108}},\ \bibinfo {pages}
  {086602} (\bibinfo {year} {2012})}\BibitemShut {NoStop}%
\bibitem [{\citenamefont {Garate}(2013)}]{garate13}%
  \BibitemOpen
  \bibfield  {author} {\bibinfo {author} {\bibfnamefont {I.}~\bibnamefont
  {Garate}},\ }\href {\doibase 10.1103/PhysRevLett.110.046402} {\bibfield
  {journal} {\bibinfo  {journal} {Phys. Rev. Lett.}\ }\textbf {\bibinfo
  {volume} {110}},\ \bibinfo {pages} {046402} (\bibinfo {year}
  {2013})}\BibitemShut {NoStop}%
\bibitem [{\citenamefont {Kane}\ and\ \citenamefont {Mele}(2005)}]{kane05}%
  \BibitemOpen
  \bibfield  {author} {\bibinfo {author} {\bibfnamefont {C.~L.}\ \bibnamefont
  {Kane}}\ and\ \bibinfo {author} {\bibfnamefont {E.~J.}\ \bibnamefont
  {Mele}},\ }\href {\doibase 10.1103/PhysRevLett.95.226801} {\bibfield
  {journal} {\bibinfo  {journal} {Phys. Rev. Lett.}\ }\textbf {\bibinfo
  {volume} {95}},\ \bibinfo {pages} {226801} (\bibinfo {year}
  {2005})}\BibitemShut {NoStop}%
\bibitem [{\citenamefont {Virtanen}\ and\ \citenamefont
  {Recher}(2012)}]{virtanen12}%
  \BibitemOpen
  \bibfield  {author} {\bibinfo {author} {\bibfnamefont {P.}~\bibnamefont
  {Virtanen}}\ and\ \bibinfo {author} {\bibfnamefont {P.}~\bibnamefont
  {Recher}},\ }\href {\doibase 10.1103/PhysRevB.85.035310} {\bibfield
  {journal} {\bibinfo  {journal} {Phys. Rev. B}\ }\textbf {\bibinfo {volume}
  {85}},\ \bibinfo {pages} {035310} (\bibinfo {year} {2012})}\BibitemShut
  {NoStop}%
\bibitem [{\citenamefont {Orth}\ \emph {et~al.}(2015)\citenamefont {Orth},
  \citenamefont {Tiwari}, \citenamefont {Meng},\ and\ \citenamefont
  {Schmidt}}]{orth15}%
  \BibitemOpen
  \bibfield  {author} {\bibinfo {author} {\bibfnamefont {C.~P.}\ \bibnamefont
  {Orth}}, \bibinfo {author} {\bibfnamefont {R.~P.}\ \bibnamefont {Tiwari}},
  \bibinfo {author} {\bibfnamefont {T.}~\bibnamefont {Meng}}, \ and\ \bibinfo
  {author} {\bibfnamefont {T.~L.}\ \bibnamefont {Schmidt}},\ }\href {\doibase
  10.1103/PhysRevB.91.081406} {\bibfield  {journal} {\bibinfo  {journal} {Phys.
  Rev. B}\ }\textbf {\bibinfo {volume} {91}},\ \bibinfo {pages} {081406(R)}
  (\bibinfo {year} {2015})}\BibitemShut {NoStop}%
\bibitem [{\citenamefont {Rod}\ \emph {et~al.}(2015)\citenamefont {Rod},
  \citenamefont {Schmidt},\ and\ \citenamefont {Rachel}}]{rod15}%
  \BibitemOpen
  \bibfield  {author} {\bibinfo {author} {\bibfnamefont {A.}~\bibnamefont
  {Rod}}, \bibinfo {author} {\bibfnamefont {T.~L.}\ \bibnamefont {Schmidt}}, \
  and\ \bibinfo {author} {\bibfnamefont {S.}~\bibnamefont {Rachel}},\ }\href
  {\doibase 10.1103/PhysRevB.91.245112} {\bibfield  {journal} {\bibinfo
  {journal} {Phys. Rev. B}\ }\textbf {\bibinfo {volume} {91}},\ \bibinfo
  {pages} {245112} (\bibinfo {year} {2015})}\BibitemShut {NoStop}%
\bibitem [{\citenamefont {Rod}\ \emph {et~al.}(2016)\citenamefont {Rod},
  \citenamefont {Dolcetto}, \citenamefont {Rachel},\ and\ \citenamefont
  {Schmidt}}]{rod16}%
  \BibitemOpen
  \bibfield  {author} {\bibinfo {author} {\bibfnamefont {A.}~\bibnamefont
  {Rod}}, \bibinfo {author} {\bibfnamefont {G.}~\bibnamefont {Dolcetto}},
  \bibinfo {author} {\bibfnamefont {S.}~\bibnamefont {Rachel}}, \ and\ \bibinfo
  {author} {\bibfnamefont {T.~L.}\ \bibnamefont {Schmidt}},\ }\href {\doibase
  10.1103/PhysRevB.94.035428} {\bibfield  {journal} {\bibinfo  {journal} {Phys.
  Rev. B}\ }\textbf {\bibinfo {volume} {94}},\ \bibinfo {pages} {035428}
  (\bibinfo {year} {2016})}\BibitemShut {NoStop}%
\bibitem [{\citenamefont {Maciejko}(2012)}]{maciejko12}%
  \BibitemOpen
  \bibfield  {author} {\bibinfo {author} {\bibfnamefont {J.}~\bibnamefont
  {Maciejko}},\ }\href {\doibase 10.1103/PhysRevB.85.245108} {\bibfield
  {journal} {\bibinfo  {journal} {Phys. Rev. B}\ }\textbf {\bibinfo {volume}
  {85}},\ \bibinfo {pages} {245108} (\bibinfo {year} {2012})}\BibitemShut
  {NoStop}%
\bibitem [{\citenamefont {Xu}\ \emph {et~al.}(2013)\citenamefont {Xu},
  \citenamefont {Yan}, \citenamefont {Zhang}, \citenamefont {Wang},
  \citenamefont {Xu}, \citenamefont {Tang}, \citenamefont {Duan},\ and\
  \citenamefont {Zhang}}]{xu13}%
  \BibitemOpen
  \bibfield  {author} {\bibinfo {author} {\bibfnamefont {Y.}~\bibnamefont
  {Xu}}, \bibinfo {author} {\bibfnamefont {B.}~\bibnamefont {Yan}}, \bibinfo
  {author} {\bibfnamefont {H.-J.}\ \bibnamefont {Zhang}}, \bibinfo {author}
  {\bibfnamefont {J.}~\bibnamefont {Wang}}, \bibinfo {author} {\bibfnamefont
  {G.}~\bibnamefont {Xu}}, \bibinfo {author} {\bibfnamefont {P.}~\bibnamefont
  {Tang}}, \bibinfo {author} {\bibfnamefont {W.}~\bibnamefont {Duan}}, \ and\
  \bibinfo {author} {\bibfnamefont {S.-C.}\ \bibnamefont {Zhang}},\ }\href
  {\doibase 10.1103/PhysRevLett.111.136804} {\bibfield  {journal} {\bibinfo
  {journal} {Phys. Rev. Lett.}\ }\textbf {\bibinfo {volume} {111}},\ \bibinfo
  {pages} {136804} (\bibinfo {year} {2013})}\BibitemShut {NoStop}%
\bibitem [{\citenamefont {Li}\ \emph {et~al.}(2015)\citenamefont {Li},
  \citenamefont {He}, \citenamefont {Meng}, \citenamefont {Xiao}, \citenamefont
  {Tang}, \citenamefont {Wei}, \citenamefont {Kim}, \citenamefont {Kioussis},
  \citenamefont {Stocks},\ and\ \citenamefont {Zhong}}]{li15c}%
  \BibitemOpen
  \bibfield  {author} {\bibinfo {author} {\bibfnamefont {J.}~\bibnamefont
  {Li}}, \bibinfo {author} {\bibfnamefont {C.}~\bibnamefont {He}}, \bibinfo
  {author} {\bibfnamefont {L.}~\bibnamefont {Meng}}, \bibinfo {author}
  {\bibfnamefont {H.}~\bibnamefont {Xiao}}, \bibinfo {author} {\bibfnamefont
  {C.}~\bibnamefont {Tang}}, \bibinfo {author} {\bibfnamefont {X.}~\bibnamefont
  {Wei}}, \bibinfo {author} {\bibfnamefont {J.}~\bibnamefont {Kim}}, \bibinfo
  {author} {\bibfnamefont {N.}~\bibnamefont {Kioussis}}, \bibinfo {author}
  {\bibfnamefont {G.~M.}\ \bibnamefont {Stocks}}, \ and\ \bibinfo {author}
  {\bibfnamefont {J.}~\bibnamefont {Zhong}},\ }\href
  {https://doi.org/10.1038%2Fsrep14115} {\bibfield  {journal} {\bibinfo
  {journal} {Sci. Rep.}\ }\textbf {\bibinfo {volume} {5}} (\bibinfo {year}
  {2015})}\BibitemShut {NoStop}%
\bibitem [{\citenamefont {Fu}\ \emph {et~al.}(2016)\citenamefont {Fu},
  \citenamefont {Ge}, \citenamefont {Su}, \citenamefont {Guo},\ and\
  \citenamefont {Liu}}]{fu16}%
  \BibitemOpen
  \bibfield  {author} {\bibinfo {author} {\bibfnamefont {B.}~\bibnamefont
  {Fu}}, \bibinfo {author} {\bibfnamefont {Y.}~\bibnamefont {Ge}}, \bibinfo
  {author} {\bibfnamefont {W.}~\bibnamefont {Su}}, \bibinfo {author}
  {\bibfnamefont {W.}~\bibnamefont {Guo}}, \ and\ \bibinfo {author}
  {\bibfnamefont {C.-C.}\ \bibnamefont {Liu}},\ }\href
  {https://doi.org/10.1038%2Fsrep30003} {\bibfield  {journal} {\bibinfo
  {journal} {Sci. Rep.}\ }\textbf {\bibinfo {volume} {6}} (\bibinfo {year}
  {2016})}\BibitemShut {NoStop}%
\bibitem [{\citenamefont {Yuan}\ \emph {et~al.}(2017)\citenamefont {Yuan},
  \citenamefont {Zhou}, \citenamefont {Wu}, \citenamefont {Zhang},
  \citenamefont {Chen}, \citenamefont {Hou},\ and\ \citenamefont
  {Wang}}]{yuan17}%
  \BibitemOpen
  \bibfield  {author} {\bibinfo {author} {\bibfnamefont {S.}~\bibnamefont
  {Yuan}}, \bibinfo {author} {\bibfnamefont {Q.}~\bibnamefont {Zhou}}, \bibinfo
  {author} {\bibfnamefont {Q.}~\bibnamefont {Wu}}, \bibinfo {author}
  {\bibfnamefont {Y.}~\bibnamefont {Zhang}}, \bibinfo {author} {\bibfnamefont
  {Q.}~\bibnamefont {Chen}}, \bibinfo {author} {\bibfnamefont {J.-M.}\
  \bibnamefont {Hou}}, \ and\ \bibinfo {author} {\bibfnamefont
  {J.}~\bibnamefont {Wang}},\ }\href
  {https://doi.org/10.1038%2Fs41699-017-0032-4} {\bibfield  {journal} {\bibinfo
   {journal} {npj 2D Materials and Applications}\ }\textbf {\bibinfo {volume}
  {1}} (\bibinfo {year} {2017})}\BibitemShut {NoStop}%
\bibitem [{\citenamefont {Garanin}\ and\ \citenamefont
  {Chudnovsky}(2015)}]{garanin15}%
  \BibitemOpen
  \bibfield  {author} {\bibinfo {author} {\bibfnamefont {D.~A.}\ \bibnamefont
  {Garanin}}\ and\ \bibinfo {author} {\bibfnamefont {E.~M.}\ \bibnamefont
  {Chudnovsky}},\ }\href {\doibase 10.1103/PhysRevB.92.024421} {\bibfield
  {journal} {\bibinfo  {journal} {Phys. Rev. B}\ }\textbf {\bibinfo {volume}
  {92}},\ \bibinfo {pages} {024421} (\bibinfo {year} {2015})}\BibitemShut
  {NoStop}%
\bibitem [{\citenamefont {Zhang}\ and\ \citenamefont {Niu}(2014)}]{zhang14a}%
  \BibitemOpen
  \bibfield  {author} {\bibinfo {author} {\bibfnamefont {L.}~\bibnamefont
  {Zhang}}\ and\ \bibinfo {author} {\bibfnamefont {Q.}~\bibnamefont {Niu}},\
  }\href {\doibase 10.1103/PhysRevLett.112.085503} {\bibfield  {journal}
  {\bibinfo  {journal} {Phys. Rev. Lett.}\ }\textbf {\bibinfo {volume} {112}},\
  \bibinfo {pages} {085503} (\bibinfo {year} {2014})}\BibitemShut {NoStop}%
\bibitem [{\citenamefont {Spencer}(2004)}]{spencer04}%
  \BibitemOpen
  \bibfield  {author} {\bibinfo {author} {\bibfnamefont {A.~J.~M.}\
  \bibnamefont {Spencer}},\ }\href@noop {} {\emph {\bibinfo {title} {Continuum
  Mechanics}}}\ (\bibinfo  {publisher} {Dover Publications},\ \bibinfo {year}
  {2004})\BibitemShut {NoStop}%
\bibitem [{\citenamefont {Maciejko}\ \emph {et~al.}(2010)\citenamefont
  {Maciejko}, \citenamefont {Qi},\ and\ \citenamefont {Zhang}}]{maciejko10}%
  \BibitemOpen
  \bibfield  {author} {\bibinfo {author} {\bibfnamefont {J.}~\bibnamefont
  {Maciejko}}, \bibinfo {author} {\bibfnamefont {X.-L.}\ \bibnamefont {Qi}}, \
  and\ \bibinfo {author} {\bibfnamefont {S.-C.}\ \bibnamefont {Zhang}},\ }\href
  {\doibase 10.1103/PhysRevB.82.155310} {\bibfield  {journal} {\bibinfo
  {journal} {Phys. Rev. B}\ }\textbf {\bibinfo {volume} {82}},\ \bibinfo
  {pages} {155310} (\bibinfo {year} {2010})}\BibitemShut {NoStop}%
\bibitem [{\citenamefont {Struck}\ and\ \citenamefont
  {Burkard}(2010)}]{struck10}%
  \BibitemOpen
  \bibfield  {author} {\bibinfo {author} {\bibfnamefont {P.~R.}\ \bibnamefont
  {Struck}}\ and\ \bibinfo {author} {\bibfnamefont {G.}~\bibnamefont
  {Burkard}},\ }\href {\doibase 10.1103/PhysRevB.82.125401} {\bibfield
  {journal} {\bibinfo  {journal} {Phys. Rev. B}\ }\textbf {\bibinfo {volume}
  {82}},\ \bibinfo {pages} {125401} (\bibinfo {year} {2010})}\BibitemShut
  {NoStop}%
\bibitem [{\citenamefont {Rudner}\ and\ \citenamefont
  {Rashba}(2010)}]{rudner10}%
  \BibitemOpen
  \bibfield  {author} {\bibinfo {author} {\bibfnamefont {M.~S.}\ \bibnamefont
  {Rudner}}\ and\ \bibinfo {author} {\bibfnamefont {E.~I.}\ \bibnamefont
  {Rashba}},\ }\href {\doibase 10.1103/physrevb.81.125426} {\bibfield
  {journal} {\bibinfo  {journal} {Phys. Rev. B}\ }\textbf {\bibinfo {volume}
  {81}},\ \bibinfo {pages} {125426} (\bibinfo {year} {2010})}\BibitemShut
  {NoStop}%
\bibitem [{\citenamefont {Ochoa}\ \emph {et~al.}(2012)\citenamefont {Ochoa},
  \citenamefont {{Castro Neto}}, \citenamefont {Fal'ko},\ and\ \citenamefont
  {Guinea}}]{ochoa12}%
  \BibitemOpen
  \bibfield  {author} {\bibinfo {author} {\bibfnamefont {H.}~\bibnamefont
  {Ochoa}}, \bibinfo {author} {\bibfnamefont {A.~H.}\ \bibnamefont {{Castro
  Neto}}}, \bibinfo {author} {\bibfnamefont {V.~I.}\ \bibnamefont {Fal'ko}}, \
  and\ \bibinfo {author} {\bibfnamefont {F.}~\bibnamefont {Guinea}},\ }\href
  {\doibase 10.1103/PhysRevB.86.245411} {\bibfield  {journal} {\bibinfo
  {journal} {Phys. Rev. B}\ }\textbf {\bibinfo {volume} {86}},\ \bibinfo
  {pages} {245411} (\bibinfo {year} {2012})}\BibitemShut {NoStop}%
\bibitem [{\citenamefont {Droth}(2014)}]{droth_phd}%
  \BibitemOpen
  \bibfield  {author} {\bibinfo {author} {\bibfnamefont {M.}~\bibnamefont
  {Droth}},\ }\emph {\bibinfo {title} {Spins and Phonons in Graphene
  Nanostructures}},\ \href
  {https://theorie.physik.uni-konstanz.de/burkard/sites/default/files/pdf/PhDThesis_Matthias-Droth.pdf}
  {Ph.D. thesis},\ \bibinfo  {school} {Universit\"at Konstanz} (\bibinfo {year}
  {2014})\BibitemShut {NoStop}%
\bibitem [{\citenamefont {Lunde}\ \emph {et~al.}(2007)\citenamefont {Lunde},
  \citenamefont {Flensberg},\ and\ \citenamefont {Glazman}}]{lunde07}%
  \BibitemOpen
  \bibfield  {author} {\bibinfo {author} {\bibfnamefont {A.~M.}\ \bibnamefont
  {Lunde}}, \bibinfo {author} {\bibfnamefont {K.}~\bibnamefont {Flensberg}}, \
  and\ \bibinfo {author} {\bibfnamefont {L.~I.}\ \bibnamefont {Glazman}},\
  }\href {\doibase 10.1103/PhysRevB.75.245418} {\bibfield  {journal} {\bibinfo
  {journal} {Phys. Rev. B}\ }\textbf {\bibinfo {volume} {75}},\ \bibinfo
  {pages} {245418} (\bibinfo {year} {2007})}\BibitemShut {NoStop}%
\bibitem [{\citenamefont {Rech}\ \emph {et~al.}(2009)\citenamefont {Rech},
  \citenamefont {Micklitz},\ and\ \citenamefont {Matveev}}]{rech09}%
  \BibitemOpen
  \bibfield  {author} {\bibinfo {author} {\bibfnamefont {J.}~\bibnamefont
  {Rech}}, \bibinfo {author} {\bibfnamefont {T.}~\bibnamefont {Micklitz}}, \
  and\ \bibinfo {author} {\bibfnamefont {K.~A.}\ \bibnamefont {Matveev}},\
  }\href {\doibase 10.1103/PhysRevLett.102.116402} {\bibfield  {journal}
  {\bibinfo  {journal} {Phys. Rev. Lett.}\ }\textbf {\bibinfo {volume} {102}},\
  \bibinfo {pages} {116402} (\bibinfo {year} {2009})}\BibitemShut {NoStop}%
\bibitem [{\citenamefont {Micklitz}\ \emph {et~al.}(2010)\citenamefont
  {Micklitz}, \citenamefont {Rech},\ and\ \citenamefont
  {Matveev}}]{micklitz10}%
  \BibitemOpen
  \bibfield  {author} {\bibinfo {author} {\bibfnamefont {T.}~\bibnamefont
  {Micklitz}}, \bibinfo {author} {\bibfnamefont {J.}~\bibnamefont {Rech}}, \
  and\ \bibinfo {author} {\bibfnamefont {K.~A.}\ \bibnamefont {Matveev}},\
  }\href {\doibase 10.1103/PhysRevB.81.115313} {\bibfield  {journal} {\bibinfo
  {journal} {Phys. Rev. B}\ }\textbf {\bibinfo {volume} {81}},\ \bibinfo
  {pages} {115313} (\bibinfo {year} {2010})}\BibitemShut {NoStop}%
\bibitem [{\citenamefont {Imambekov}\ \emph {et~al.}(2012)\citenamefont
  {Imambekov}, \citenamefont {Schmidt},\ and\ \citenamefont
  {Glazman}}]{imambekov12}%
  \BibitemOpen
  \bibfield  {author} {\bibinfo {author} {\bibfnamefont {A.}~\bibnamefont
  {Imambekov}}, \bibinfo {author} {\bibfnamefont {T.~L.}\ \bibnamefont
  {Schmidt}}, \ and\ \bibinfo {author} {\bibfnamefont {L.~I.}\ \bibnamefont
  {Glazman}},\ }\href {\doibase 10.1103/RevModPhys.84.1253} {\bibfield
  {journal} {\bibinfo  {journal} {Rev. Mod. Phys.}\ }\textbf {\bibinfo {volume}
  {84}},\ \bibinfo {pages} {1253} (\bibinfo {year} {2012})}\BibitemShut
  {NoStop}%
\bibitem [{\citenamefont {Schmidt}(2013)}]{schmidt13b}%
  \BibitemOpen
  \bibfield  {author} {\bibinfo {author} {\bibfnamefont {T.~L.}\ \bibnamefont
  {Schmidt}},\ }\href {\doibase 10.1103/PhysRevB.88.235429} {\bibfield
  {journal} {\bibinfo  {journal} {Phys. Rev. B}\ }\textbf {\bibinfo {volume}
  {88}},\ \bibinfo {pages} {235429} (\bibinfo {year} {2013})}\BibitemShut
  {NoStop}%
\bibitem [{\citenamefont {Ziman}(2001)}]{ziman01}%
  \BibitemOpen
  \bibfield  {author} {\bibinfo {author} {\bibfnamefont {J.~M.}\ \bibnamefont
  {Ziman}},\ }\href@noop {} {\emph {\bibinfo {title} {Electrons and Phonons:
  The Theory of Transport Phenomena in Solids}}}\ (\bibinfo  {publisher}
  {Oxford University Press},\ \bibinfo {year} {2001})\BibitemShut {NoStop}%
\bibitem [{\citenamefont {Martienssen}\ and\ \citenamefont
  {Warlimont}(2005)}]{handbook05}%
  \BibitemOpen
  \bibinfo {editor} {\bibfnamefont {W.}~\bibnamefont {Martienssen}}\ and\
  \bibinfo {editor} {\bibfnamefont {H.}~\bibnamefont {Warlimont}},\ eds.,\
  \href@noop {} {\emph {\bibinfo {title} {Springer Handbook of Condensed Matter
  and Materials Data}}}\ (\bibinfo  {publisher} {Springer-Verlag Berlin
  Heidelberg},\ \bibinfo {year} {2005})\BibitemShut {NoStop}%
\bibitem [{\citenamefont {Bid}\ \emph {et~al.}(2006)\citenamefont {Bid},
  \citenamefont {Bora},\ and\ \citenamefont {Raychaudhuri}}]{aveek06}%
  \BibitemOpen
  \bibfield  {author} {\bibinfo {author} {\bibfnamefont {A.}~\bibnamefont
  {Bid}}, \bibinfo {author} {\bibfnamefont {A.}~\bibnamefont {Bora}}, \ and\
  \bibinfo {author} {\bibfnamefont {A.~K.}\ \bibnamefont {Raychaudhuri}},\
  }\href {\doibase 10.1103/physrevb.74.035426} {\bibfield  {journal} {\bibinfo
  {journal} {Phys. Rev. B}\ }\textbf {\bibinfo {volume} {74}},\ \bibinfo
  {pages} {035426} (\bibinfo {year} {2006})}\BibitemShut {NoStop}%
\end{thebibliography}%

\end{document}